\documentclass[final,5p,times,twocolumn]{elsarticle}
\usepackage{graphicx}
\usepackage{amsmath,amssymb,bm,amsfonts}
\usepackage{citesort}
\usepackage{slashed}
\newcommand{\Lagr}{\mathcal{L}}
\newcommand{\reff}[1]{(\ref{#1})}
\renewcommand{\Re}{\operatorname{Re}}
\renewcommand{\Im}{\operatorname{Im}}
\allowdisplaybreaks[1]
\begin{document}

\begin{frontmatter}
\title{New insights into antikaon-nucleon scattering and the
    structure of the $\Lambda (1405)$}

\author[Bonn]{Maxim Mai}
\author[Bonn,Julich]{Ulf-G. Mei{\ss}ner}

\address[Bonn]{Universit\"at Bonn,
             Helmholtz-Institut f\"ur Strahlen- und Kernphysik (Theorie)
         and Bethe Center for Theoretical Physics,
             D-53115 Bonn, Germany}

\address[Julich]{Forschungszentrum J\"ulich, 
             Institut f\"ur Kernphysik,
             Institute for Advanced Simulation, 
         and J\"ulich Center for Hadron Physics,
             D-52425 J\"ulich, Germany}
\date{\today}

\begin{abstract}
We perform a combined analysis of antikaon-nucleon scattering cross
sections and the recent SIDDHARTA kaonic hydrogen data in the framework
of a coupled-channel Bethe-Salpeter approach at next-to-leading order
in the chiral expansion of the effective potential.
We find a precise description of the antikaon-proton scattering amplitudes 
and are able to extract accurate values of the scattering lengths,
$a_0  =-1.81^{+0.30}_{-0.28} + i~ 0.92^{+0.29}_{-0.23}~{\rm fm}$~,
$a_1  =+0.48^{+0.12}_{-0.11} + i~ 0.87^{+0.26}_{-0.20}~{\rm fm}$.
We also discuss the two-pole structure of the $\Lambda(1405)$. 
\end{abstract}


\end{frontmatter}

\section{Introduction and summary}
With the recent precise measurement of the characteristics 
of kaonic hydrogen by the SIDDHARTA collaboration \cite{Bazzi:2011zj},
an accurate determination of the so important antikaon-nucleon scattering
amplitude is now possible. The appropriate framework to perform this task is
unitarized chiral perturbation theory, which combines the strictures
from the chiral SU(3) dynamics of QCD with  coupled channel effects,
that e.g. generate the much discussed $\Lambda (1405)$ resonance, as
first pointed out by Dalitz and Tuan \cite{Dalitz:1960du}. From 
earlier studies by various groups, it is already known that simply
taking the leading order chiral interactions in the effective potential
of the respective scattering equation is insufficient to achieve the
desired accurate theoretical description, 
see e.g. Refs.~\cite{Borasoy:2005ie,Oller:2006jw,Borasoy:2006sr}.
In fact, Ikeda et al. \cite{Ikeda:2011pi,Ikeda:2012au}
have performed such a combined analysis based
on the next-to-leading order chiral effective meson-baryon Lagrangian,
nicely demonstrating that indeed a more precise description of the
$K^-p$ and $K^-n$ interaction arises. Here, we perform a similar
analysis, but in contrast to Refs.~\cite{Ikeda:2011pi,Ikeda:2012au},
we use a Bethe-Salpeter framework without an on-shell approximation
for the intermediate meson-baryon states (as described in more detail
below). The framework we use has already been successfully applied
to pion-nucleon scattering in the s-waves \cite{Bruns:2010sv} and
thus it is evident to extend this analysis to antikaon-nucleon scattering.
We also point out that the constraints from SIDDHARTA on the kaon-deuteron
scattering length have been investigated in \cite{Doring:2011xc}.

The main results of our investigation can be summarized as follows:
\begin{itemize}
\item Fitting the scattering data for $K^-p \to K^-p$, $\bar K^0n$,
$\Sigma^\pm\pi^\mp$, and $\Sigma^0\pi^0$ for laboratory momenta
$p_{\rm lab} \leq 300\,$MeV together with the SIDDHARTA data allows
for a good description of the antikaon-proton cross section data 
(cf. Fig.~\ref{pic:WQ}) and an accurate determination of the 
scattering lengths,  cf. Eq.~(\ref{eq:a}).
\item We can give a precise prediction for the real and imaginary part
of the $K^-p \to K^-p$ scattering amplitude for center-of-mass energies
$1330\,{\rm MeV} \leq W_{\rm cms} \leq 1450\,$MeV, cf. Fig.~\ref{pic:PWA}.
\item We have investigated the two-pole structure of the 
$\Lambda(1405)$ \cite{Oller:2000fj,Jido:2003cb}.
While the first pole is in agreement with other determinations, 
we find the real part of the second pole at larger energies than 
usually obtained. We trace this back to
the fitting procedures in other works that restrict the NLO
contributions to come out
close to the solution given by the Weinberg-Tomozawa term, while 
we do not impose such a restriction. 
\end{itemize}

\section{Framework}
The starting point of our work is the recent analysis of $\pi N$ scattering, 
presented in \cite{Bruns:2010sv}. In this section we describe the basic ingredients
of this approach. 

In chiral perturbation theory the meson-baryon interaction at the leading 
chiral order is encoded in the following Lagrangian
	\begin{align}
	\Lagr^{(1)}_{\phi B}&=\langle \bar{B} (i\gamma_\mu D^\mu-m_0)B\rangle
	+\frac{D/F}{2}\langle \bar{B}\gamma_\mu \gamma_5[u^\mu,B]_{+/-} \rangle ~,
	\end{align}
where $\langle\ldots\rangle$ denotes the trace in flavor space, $D_\mu B
:=\partial_\mu B 
+\frac{1}{2}[[u^\dagger,\partial_\mu u],B]$, $m_0$ is the baryon octet mass in
the chiral SU(3) limit, and $D,F$ are the axial coupling constants. The
relevant degrees of freedom are the Goldstone bosons described by the
traceless meson matrix $\phi$, which is included in the above Lagrangian 
via $u^2:=\exp\bigl(i\phi/F_0\bigr)$ and $u^\mu:=iu^{\dagger}\partial^\mu u 
- iu\partial^\mu u^{\dagger}$. Here, $F_0$ is the meson decay constant in the 
chiral limit. The low-lying baryons are collected in the traceless matrix
$B$. We set the external currents to zero except for the scalar one, which is 
set equal to the quark mass matrix, i.e. $s=\mathcal{M}:=\textrm{diag}(m_u,
m_d, m_s)$.  We furthermore use $\chi_\pm:=u^{\dagger}\chi u^{\dagger}\pm u 
\chi^{\dagger}u$ and $\chi:=2B_0 s$, where the constant $B_0$ is related to 
the quark condensate in the chiral limit.

Starting from the covariant derivative $D_\mu B$, the so-called
Weinberg-Tomozawa term can be derived. This term dominates the s-wave 
interaction near the thresholds, therefore in most chiral unitary approaches 
the meson-baryon interaction is restricted to this term. Secondly, a meson 
can couple to a baryon via the axial vector current $\sim D,F$, generating 
the $s$- and $u$-channel exchanges of the intermediate baryons. The inclusion 
of these so-called Born graphs in the driving term of the Bethe-Salpeter
equation leads to conceptional and practical difficulties, which are described 
in detail in Ref.~\cite{Bruns:2010sv, Mai:2011xx}. The latter are usually
overcome, making use of the on-shell approximation or via projection of the 
kernel to the s-wave, see e.g. Ref.~\cite{Borasoy:2006sr} and 
Ref.~\cite{Ikeda:2011pi} for a more recent study. However the particular
attention of the present work lies on the solution of the Bethe-Salper 
equation with the full off-shell dependence. Thus we will restrict the
interaction kernel to a sum of contact terms, but refrain from the 
approximations mentioned above.

Aside from the Weinberg-Tomozawa term, we will take into account the full set
of meson-baryon vertices from the second-order chiral Lagrangian. The
pertinent Lagrangian density was first constructed in \cite{Krause:1990xc} 
and reads in its minimal form~\cite{Frink:2004ic}
	\begin{align}\label{eqn:LAGR}
	&\Lagr^{(2)}_{\phi B}= b_{D/F} \langle\overline B
	\big[\chi_+,B\big]_\pm\rangle
	+b_0 \langle\overline B B\rangle \langle\chi_+\rangle\nonumber\\
	&+b_{1/2} \langle\overline B  \Big[u_\mu,\big[u^\mu,B\big]_\mp\Big]\rangle
	+b_3 \langle\overline B \Big\{ u_\mu,\big\{ u^\mu,B\big\}\Big\}\rangle
	+b_4 \langle\overline B  B\rangle \langle u_\mu u^\mu\rangle \nonumber\\
	&+ib_{5/6}  \langle\overline B\sigma^{\mu\nu} \Big[\big[u_\mu,u_\nu\big], B\Big]_\mp\rangle
	+ib_7 \langle\overline B\sigma^{\mu\nu} u_\mu\rangle  \langle u_\nu B\rangle \nonumber\\
	&+ \frac{i\,b_{8/9}}{2m_0}\Big( \langle\overline B \gamma^\mu\Big[u_\mu,\big[u_\nu,\big[D^\nu, B\big]\big]_\mp\Big]\rangle+\langle\overline B \gamma^\mu\Big[D_\nu,\big[u^\nu,
	\big[u_\mu,B\big]\big]_\mp\Big]\rangle\Big) \nonumber\\
	&+\frac{i\,b_{10}}{2m_0}\Big( \langle\overline B 
	\gamma^\mu\Big\{ u_\mu,\big\{ u_\nu,\big[D^\nu,B\big]\big\}\Big\}\rangle+\langle
	\overline B\gamma^\mu\Big[D_\nu,\big\{ u^\nu,
	\big\{ u_\mu,B\big\}\big\}\Big]\rangle\Big) \nonumber\\
	&+\frac{i\,b_{11}}{2m_0}\Big( 2\langle\overline B \gamma^\mu 
	\big[D_\nu,B\big]\rangle \langle u_\mu u^\nu\rangle\nonumber\\
	&\qquad\qquad\quad +\langle\overline B \gamma^\mu B\rangle 
	\langle\big[D_\nu,u_\mu\big]u^\nu + u_\mu \big[D_\nu,u^\nu\big]\rangle   \Big)~,
	\end{align}
with the $b_i$ the pertinent dimension-two low energy constants (LECs). The
LECs $b_{0,D,F}$ are the so-called {\it symmetry breakers}, while the $b_i$
$(i= 1,\ldots, 11)$ are referred to as {\it dynamical} LECs. On the one hand
such terms may lead to sizable corrections to the leading-order result, see 
e.g. \cite{Mai:2009ce} for the calculation of meson-baryon scattering lengths 
up to the third chiral order. On the other hand, including such terms with 
full off-shell dependence we hope to account for some of the structures created 
by the missing Born graphs. Let us denote the in- and out-going meson momenta by $q_1$
and $q_2$, respectively. The overall four-momentum is given by $p=q_1+p_1=q_2+p_2$, 
where $p_1$ and $p_2$ are the momenta of in- and out-going baryon,
respectively. Separating the momentum space from the channel
space structures, the chiral potential considered here takes the form:
	\begin{align*}
	V(\slashed{q}_2, \slashed{q}_1; p)&=A_{WT}(\slashed{q_1}+\slashed{q_2})\\
	&\quad+A_{14}(q_1\cdot q_2)+A_{57}[\slashed{q_1},\slashed{q_2}]+A_{M}\\
	&\quad+A_{811}\Big(\slashed{q_2}(q_1\cdot p)+\slashed{q_1}(q_2\cdot p)\Big),
	\end{align*} 
where the first matrix $A_{WT}$ only depends on the meson decay constants $F_{\pi,K}$, 
whereas  $A_{14}$, $A_{57}$, $A_{811}$ and $A_{M}$ also contain the NLO LECs
as specified in appendix  A. In going from the
Lagrangian Eq.~\reff{eqn:LAGR} to the above vertex rule, we have left out some 
terms, which are formally of third chiral order. 
The channel space is defined in accordance with the quantum numbers as well 
as the energy range of interest. For the purpose of gaining some insight on 
the nature of $\Lambda(1405)$ it is spanned by six vectors corresponding to 
the following meson-baryon states: $\{K^-p;~\bar K^0 n;~\pi^0\Lambda;~\pi^0\Sigma^0;
~\pi^+\Sigma^-;~\pi^-\Sigma^+\}$. The influence of other much heavier channels 
will be absorbed in the LECs to be fitted.

The strict perturbative chiral expansion is only applicable at low energies
and certainly fails in the vicinity of (subthreshold) resonances. We extend the 
range of applicability by means of a coupled channel Bethe-Salpeter equation
(BSE). Introduced in Ref.~\cite{Salpeter:1951sz} it has been proven to be very 
useful both in the purely mesonic and also in the meson-baryon  sector. In
contrast to perturbative calculations this approach implements two-body
unitarity exactly and in principle allows to generate resonances dynamically. 
For the meson-baryon scattering amplitude $T(\slashed{q}_2, 
\slashed{q}_1; p)$ and the chiral potential $V(\slashed{q}_2, \slashed{q}_1;
p)$ the integral equation to be solved reads
	\begin{align}\label{eqn:BSE}
	T(\slashed{q}_2, \slashed{q}_1; p)= &V(\slashed{q}_2, \slashed{q}_1;
        p) +\nonumber\\
	&i\int\frac{d^d l}{(2\pi)^d}V(\slashed{q}_2, \slashed{l}; p) 
         S(\slashed{p}-\slashed{l})\Delta(l)T(\slashed{l}, \slashed{q}_1; p),
	\end{align}
where $S$ and $\Delta$ represent the baryon (of mass $m$) and the meson (of
mass $M$) propagator, respectively, and are given by $iS(\slashed{p}) =
{i}/({\slashed{p}-m+i\epsilon})$ and $i\Delta(k) ={i}/({k^2-M^2+i\epsilon})$. 
Moreover, $T$, $V$, $S$ and $\Delta$ in the last expression are matrices in 
the channel space.

To treat the loop diagrams appearing in the BSE Eq.~\reff{eqn:BSE}, we utilize 
dimensional regularization. The purely baryonic integrals are set to zero 
from the beginning. In the spirit of our previous work \cite{Bruns:2010sv},
we apply the usual $\overline{\rm MS}$ subtraction scheme, keeping in mind 
that the modified loop integrals are still scale-dependent. The scale $\mu$ 
reflects the influence of the higher-order terms not included in our
potential. It is used as a fitting parameter of our approach.

The solution of the BSE Eq.~\reff{eqn:BSE} with full off-shell dependence is 
obtained following the construction principles described in 
Ref.~\cite{Bruns:2010sv}. As an extension of this approach we wish also 
to address an other issue here, namely analyticity. Let us first start 
with the one-meson-one-baryon loop-function $I_{MB}(s=p^2)$ in four dimensions.
	\begin{align*}
	 I_{MB}(s):=\int\frac{d^4 l}{(2\pi)^4}\frac{1}{l^2-M^2}\frac{1}{(p-l)^2-m^2}
	\end{align*}
Applying the Cutkosky rules, one immediately obtains the imaginary part of 
this integral, given by $-({q_{\rm cms}(p^2))}/({8\pi\sqrt{s}})$ for 
$q_{\rm cms}^2=\left((s-(m+M)^2)(s-(m-M)^2)\right)/(4s)$. Keeping in mind the 
high energy behavior of the function $I_{MB}(s)$, we can obtain the real 
part of it via a subtracted dispersion relation
	\begin{align*}
	 {\rm Re} \big(I_{MB}(s)\big)= {\rm Re} \big(I_{MB}(s_0)\big)+
         \frac{(s-s_0)}{\pi}\int_{s_{\rm thr}}^{\infty}ds'
          \frac{\Im\big(I_{MB}(s')\big)}{(s-s')(s-s_0)},
	\end{align*}
where $s_{\rm thr}=(M+m)^2$ and $s_0$ is a subtraction point chosen to not lie 
on the integration contour. The solution of the BSE Eq.~\reff{eqn:BSE} corresponds 
to a bubble sum, containing exactly the same one loop-functions $I_{MB}$. Thus 
one should in principle be able to write an equation similar to the last one 
for the scattering amplitude
	\begin{align}\label{eqn:DISP}
	 {\rm Re} \big(T(s)\big)= {\rm Re} \big(T(s_0)\big) 
         +\frac{(s-s_0)}{\pi}\int_{s_{\rm thr}}^{\infty}ds'\frac{\Im\big(T(s')\big)}{(s-s')(s-s_0)},
	\end{align}
where for the moment we have suppressed the $\slashed{q}_{1,2}$ dependence. 
To the best of our knowledge, it is not possible to implement 
the last relation Eq.~\reff{eqn:DISP} into the BSE Eq.~\reff{eqn:BSE} directly. 
To put it in other words, the BSE ansatz is known to produce poles on the physical
Riemann sheet, which are forbidden by the postulate of maximal analyticity. Thus a scattering amplitude, which solves the Eq.~\reff{eqn:BSE}, does not satisfy Eq.~\reff{eqn:DISP}.

Nevertheless it is possible to find a solution of the BSE Eq.~\reff{eqn:BSE},
which fulfills Eq.~\reff{eqn:DISP} at least approximately, as we wish to describe now. One way to do so is to
keep only those solutions of the BSE, which do not produce poles on the first Riemann
sheet 'near' the real (physical) axis. E.g. in Ref.~\cite{Borasoy:2006sr} solutions producing poles for
$\Im(W_{\rm cms}=\sqrt{s})<250\,$MeV were excluded by hand. To overcome such unsatisfactory
intervention into the fitting procedure we proceed differently. First, for a fixed $s$
and $s_0$ we define the following quantity
	\begin{align}\label{eqn:DISP1}
	 \chi^2_{\rm DISP} = \left(\frac{ {\rm Re}
             \left(T(s)-T(s_0)\right)-\frac{(s-s_0)}{\pi}
             \int_{s_{\rm thr}}^{\infty}ds'\frac{\Im\big(T(s')\big)}{(s-s')(s-s_0)}}
              {{\rm Re}\left(T(s)-T(s_0)\right)}\right)^2.
	\end{align}
Then the fitting parameters of our model are adjusted to minimize the 
quantity $\chi^2_{\rm FULL}=\chi^2_{\rm DISP}+\chi^2_{\rm DATA}$, where the
latter is based on the experimental data. It should be clear that such a
procedure is not suited to overcome the unphysical poles. It ensures, however,
that they are moved far away from the real axis in a systematic manner, without
manual intervention. This we consider an improvement of the model.

\section{Fit strategy}
We are now able to confront our approach with the experimental results. 
Throughout the present work we use the following numerical values (in GeV) 
for the masses and the meson decay constants:
 $F_\pi=0.0924$, $~F_K =0.113$, $M_{\pi^0}=0.135$, $M_{\pi^{+/-}}=0.1396$, 
 $M_{K^-}=0.4937$, $M_{\bar K^0}=0.4977$, $m_p=0.9383$,
 $m_n=0.9396$, $m_\Lambda=1.1157$, $m_{\Sigma^0}=1.1926$,
 $m_{\Sigma^+}=1.1894$ and $m_{\Sigma^-}=1.1975$. 
The baryon mass in the chiral limit, $m_0$ in Eq.~\reff{eqn:LAGR}, can be
fixed to $1$~GeV without loss of generality, as any other value only amounts 
to a rescaling  of the unknown LECs.

Secondly, for the experimental data we consider total cross sections for 
the processes $K^-p\to\{K^-p,~\bar K^0 n,~\pi^0\Sigma^0,~\pi^+\Sigma^-,
~\pi^-\Sigma^+\}$ taken from 
Refs.~\cite{Ciborowski:1982et, Humphrey:1962zz, Sakitt:1965kh,
  Watson:1963zz}. 
Moreover, we consider the following decay ratios
	\begin{align*}
	\gamma&=\frac{\Gamma(K^-p\rightarrow \pi^+\Sigma^-)}{\Gamma(K^-p\rightarrow \pi^-\Sigma^+)}=2.38\pm0.04,\\
	R_n&=\frac{\Gamma(K^-p\rightarrow \pi^0\Lambda)}{\Gamma(K^-p\rightarrow \text{neutral states})}=0.189\pm0.015,\\
	R_c&=\frac{\Gamma(K^-p\rightarrow \pi^+\Sigma^-, \pi^-\Sigma^+)}{\Gamma(K^-p\rightarrow \text{inelastic channels})}=0.664\pm0.011,
	\end{align*}
where the first one is taken from Ref.~\cite{Tovee:1971ga} and the last 
two from Ref.~\cite{Nowak:1978au}. Additionally to these quite old date
we use a recent determination of the energy shift and width of the kaonic
hydrogen in the 1s state, i.e. $\Delta E -i\Gamma/2=(283\pm42)-i(271\pm55)$ eV
from the SIDDHARTA experiment at DA$\Phi$NE \cite{Bazzi:2011zj}. These 
are related to the $K^-p$ scattering length via the modified Deser-type
relation \cite{Meissner:2004jr}
	\begin{align*}
	\Delta E -i\Gamma/2=-2\alpha^3\mu^2_ca_{K^-p}
        \left[1-2a_{K^-p}\alpha\mu_c(\ln \alpha -1)\right],
	\end{align*}
where $\alpha \simeq 1/137$ is the  fine-structure constant, $\mu_c$ is the reduced 
mass and $a_{K^-p}$ the scattering length of the $K^-p$ system.

There are 17 free parameters in the present approach. First of all, 
the low-energy constants represent the heavy degrees of freedom of QCD, 
which are integrated out. Thus they have to be fixed in a fit to the 
experimental data. As a matter of fact, the fitting parameters of our approach 
correspond to the SU(3) low-energy constants, renormalized by the effects of 
the not included channels $\{\eta\Lambda; \eta \Sigma^0; K^+\Xi^-;
K^0\Xi^0\}$. Additionally, three subtraction constants have to be determined 
from a fit, which correspond to the logarithms of the undetermined
regularization scales $\{\mu_{\pi\Lambda}$, $\mu_{KN}$, $\mu_{\pi\Sigma}\}$.

To reproduce the experimental data as well as to preserve the property of analyticity
as described in above, we minimize the following quantity 
$\chi_{\rm FULL}^2=\chi_{\rm  DISP}^2+\chi^2_{\rm DATA}$, where the first part is given in the 
Eq.~\reff{eqn:DISP1} and the second part by the quantity
	\begin{align*}
	\chi^2_{\rm DATA}:=\frac{\chi^2}{\rm d.o.f}
         =\frac{\sum_i n_i}{N(\sum_i n_i-p)}\sum_i\frac{\chi^2_i}{n_i}.
	\end{align*}
Here $p$ is the numbers of the free parameters, $n_i$ is the number of data
points available for the observable $i$ and $N$ is the number of observables.
The present choice of $\chi^2_{\rm DATA}$ is crucial to ensure
the equal weight of different observables, independently of the corresponding 
number of data points. The minimization itself is performed using MINUIT \cite{MINUIT},
especially the migrad strategy in two steps, which is due to the quite
complicated structure of the BSE solution with full off-shell dependence. 
First, parameters are found to minimize the $\chi^2_{\rm FULL}$ in the
on-shell parametrization. In the second step we turn on the ``off-shellness'' 
slowly, minimizing in each step the $\chi^2_{\rm DATA}$ and taking the
parameters of the best fit from the previous step as starting values. Such 
a procedure guarantees preservation of the right analytic properties of the 
solution, found in the first step.

\vfill

\section{Results}

	\begin{figure*}[thb]
	\includegraphics[keepaspectratio = false,width=1.0\textwidth, height=0.6\textheight]{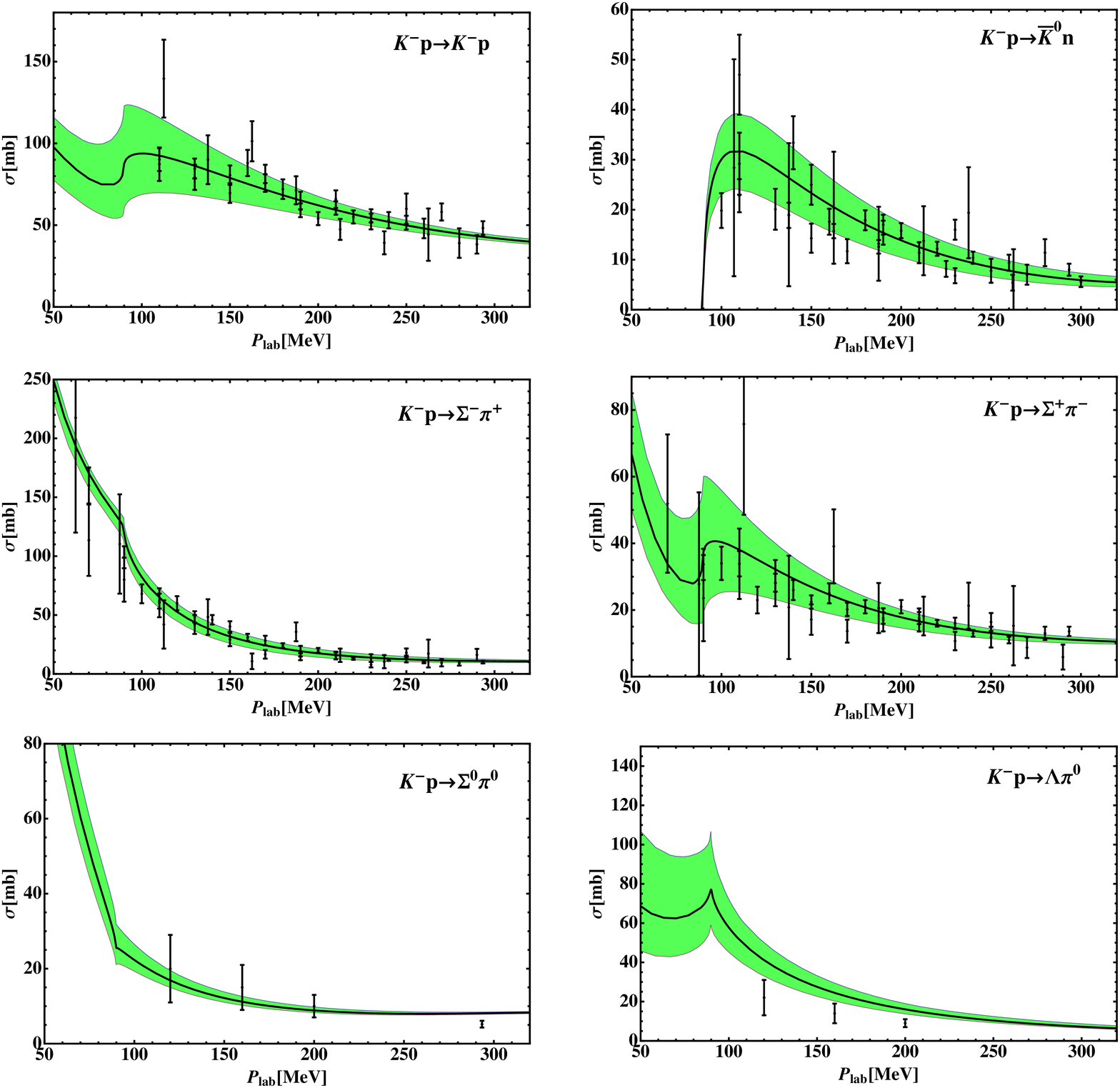}
	\caption{Total cross sections for the scattering of $K^-p$ to various
         channels versus  the $K^-$ laboratory momentum.
         The black points with error 
	 bars denote the experimental data from \cite{Ciborowski:1982et, Humphrey:1962zz, 
	Sakitt:1965kh, Watson:1963zz} considered for the fits. Shaded (green)
        bands denote the $1\sigma$ 
	error bands calculated as described in the text. The reaction $K^-p \to \Lambda\pi^0$ is 
	not a part of our fit and presented here only for completeness.}\label{pic:WQ}
	\end{figure*}

For the best fit $\tilde\chi^2_{DATA}=0.524$ we obtain the following parameter 
set (all $b_i$ in GeV$^{-1}$ and $\mu_{i}$ in GeV)
\begin{center}
\begin{tabular}{c c c c}
\hline
~~~&$\log(\mu_{KN}/(1{\rm GeV}))		$&$=+1.155\pm0.181$&~~~\\
~~~&$\log(\mu_{\pi\Sigma}/(1{\rm GeV})) 	$&$=-0.008\pm0.002$&~~~\\
~~~&$\log(\mu_{\pi\Lambda}/(1{\rm GeV}))	$&$=-0.010\pm0.003$&~~~\\
\end{tabular}
\begin{tabular}{c | c}
\hline
$b_{1~}=+0.582\pm0.052$		&$b_{8~}=-0.332\pm0.045$\\
$b_{2~}=-0.310\pm0.092$		&$b_{9~}=+0.298\pm0.087$\\
$b_{3~}=+0.227\pm0.038$		&$b_{10}=+0.198\pm0.058$\\
$b_{4~}=-0.939\pm0.069$		&$b_{11}=+0.516\pm0.058$\\
$b_{5~}=+0.023\pm0.007$		&$b_{0~}=+0.710\pm0.211$\\
$b_{6~}=+0.001\pm0.001$		&$b_{D~}=-0.291\pm0.068$\\
$b_{7~}=-2.518\pm0.110$		&$b_{F~}=-0.057\pm0.014$\\
\hline
\end{tabular}
\end{center}

We note that the LECs are all of natural size, indicating that all relevant
physical mechanisms are included in the calculations. The experimental data on
total cross sections is reproduced quite nicely, see Fig.\ref{pic:WQ}. We first 
wish to remark that due to insufficient number of data points the channel $K^-p\to\Lambda\pi^0$
is not considered as experimental input in the fit procedure. For completeness, we present the
outcome of our approach for this channel in Fig.\ref{pic:WQ}. Secondly, all of the 
cross sections presented here are due to the strong interaction only. Additionally, Coulomb
interaction was taken into account in \cite{Borasoy:2006sr, Ikeda:2012au} via a non-relativistic
quantum mechanical formula. Since this alone cannot count for an interference between
the strong and the electromagnetic interactions, we relegate the proper
inclusion of the electromagnetic contributions to a future publication.

The confidence bands presented in the above figure and the errors on further observables
are calculated as follows: first we generate a large number
($\sim10,000$) of randomly distributed parameter sets in the error region 
given above. Then for each of these parameter sets we calculate the
$\chi^2_{\rm DATA}$ and keep only those sets, for which $\chi^2_{\rm
  DATA}-\tilde\chi^2_{\rm DATA}\leq 1.05$.
Quantities calculated for these parameter sets are assumed to lie in the
$1\sigma$ region around the central value.

\begin{figure}[t]
	\begin{center}
	\includegraphics[width=0.95\linewidth]{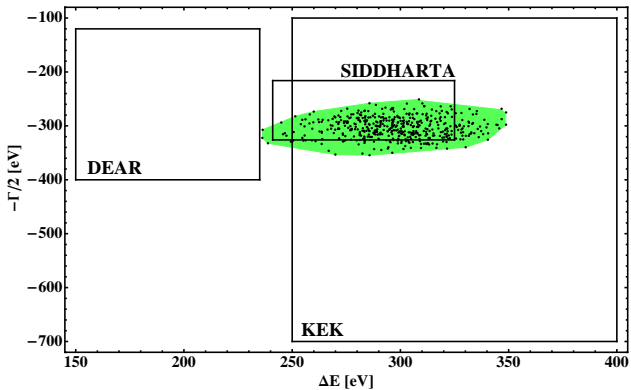}
	\end{center}
	\caption{Energy shift and width of  kaonic hydrogen as determined
          from the DEAR \cite{Beer:2005qi}, the KEK \cite{Ito:1998yi} and the 
          SIDDHARTA \cite{Bazzi:2011zj} experiments. The shaded area denotes 
          the $1\sigma$ region of our approach around the best fit value.}
         \label{pic:SIDD}
	\end{figure}

\begin{figure}[t]
	\begin{center}
	\includegraphics[width=1.0\linewidth]{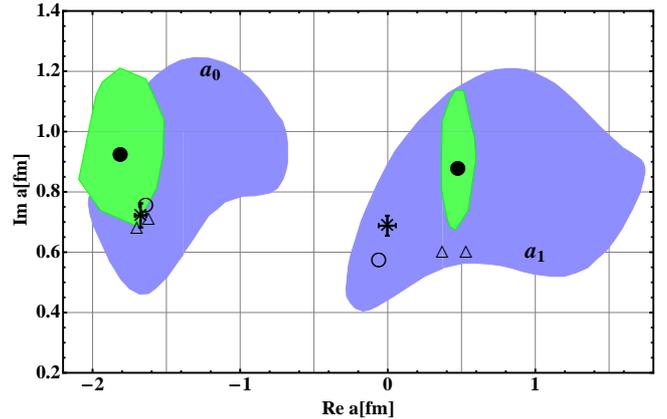}
	\end{center}
	\caption{Real and imaginary part of isospin 0 and 1 $KN\to KN$
          scattering lengths. The light shaded (green) areas correspond to 
          the $1\sigma$ region of our approach around the central value 
          (full circles). The darker (blue) areas correspond to the $1\sigma$ 
          region around central value (empty circle) from
          Ref.~\cite{Borasoy:2006sr}. 
          the cross and empty triangles denote older experimental values 
          from \cite{Kim:1965zz} and \cite{Martin:1980qe}, respectively.}
         \label{pic:a0}
	\end{figure}

The results for the threshold quantities are in excellent agreement with experimental data and read
	\begin{align*}
	\Delta E -i\Gamma/2&=+296^{+56}_{-49}-i~300^{+42}_{-54}~\text{eV},\\
	\gamma&=+2.44^{+0.73}_{-0.67}~,\\
	R_n&=+0.268^{+0.110}_{-0.086}~,\\
	R_c&=+0.643^{+0.015}_{-0.019}~.
	\end{align*}
As a matter of fact, the shape of the $1\sigma$ region for the energy 
shift and width of kaonic hydrogen cannot be assumed to be rectangular, 
see Fig.~\ref{pic:SIDD}. The resulting scattering lengths for  isospin
$I=0$ and $I=1$, i.e. $a_0$ and $a_1$, are displayed in Fig.~\ref{pic:a0}, in comparison
to some older determinations and the determination based on scattering data alone
\cite{Borasoy:2006sr}. The inclusion of the SIDDHARTA data leads to much smaller
errors, especially for $a_1$. Our values for the scattering lengths are 
\begin{align}\label{eq:a}
a_0    &=-1.81^{+0.30}_{-0.28} + i~ 0.92^{+0.29}_{-0.23}~{\rm fm}~,\nonumber\\
a_1    &=+0.48^{+0.12}_{-0.11} + i~ 0.87^{+0.26}_{-0.20}~{\rm fm}~.
\end{align}
The inclusion of the $\Lambda\pi^0$ data in the fitting procedure 
could yield an additional constraint on the isospin $I=1$ amplitudes 
and fix the value of $a_1$ as done in \cite{Borasoy:2006sr}. We have
not considered this channel as an experimental input for the reasons given above.
The scattering length for the elastic $K^-p$ channel reads 
$a_{K^-p} =-0.68^{+0.18}_{-0.17} + i~ 0.90^{+0.13}_{-0.13}~{\rm fm}$.
For comparison, taking the SIDDHARTA data only, one obtains
$a_{K^-p}=-0.65^{+0.15}_{-0.15} + i~ 0.81^{+0.18}_{-0.18}$ fm, while Ikeda et al. 
find\footnote{Here, the error bars are extracted from Fig. 4 of
\cite{Ikeda:2012au}.} $a_{K^-p}=-0.70^{+0.13}_{-0.13} + i~ 0.89^{+0.16}_{-0.16}$ fm. Therefore, 
these fundamental chiral SU(3) parameters can now be
considered to be determined with about an accuracy of $\sim 15\%$,

\begin{figure*}[t]
	\includegraphics[width=1.0\textwidth]{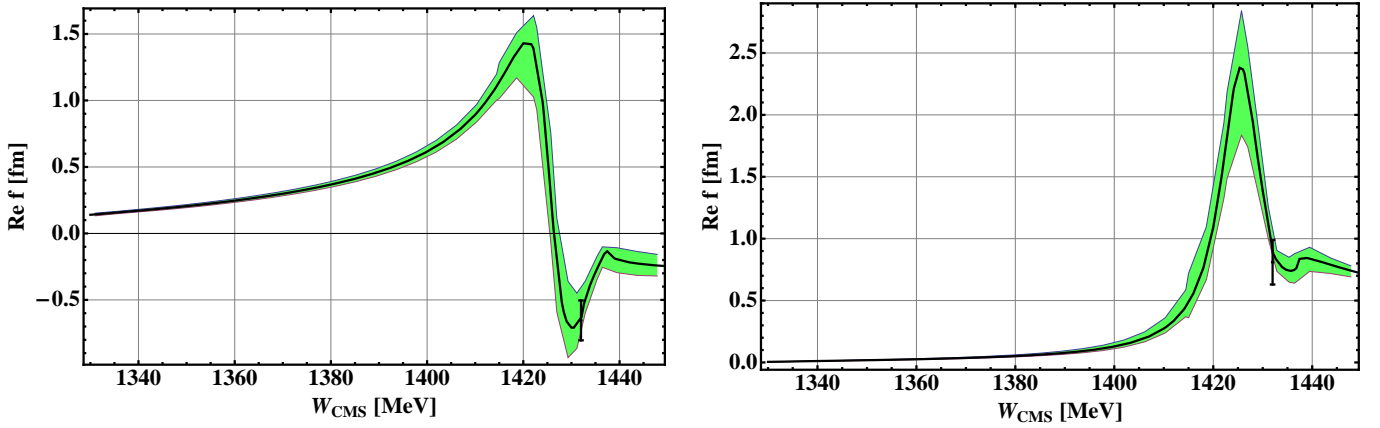}
	\caption{Real and imaginary part of the $K^-p\to K^-p$ scattering 
                 amplitude. The shaded band indicates the uncertainty of the 
                 calculation. The data point at $W_{\rm cms}=M_K+m_p$ is
                 determined from the energy shift and width of kaonic hydrogen 
                 from the SIDDHARTA experiment.}
                 \label{pic:PWA}
	\end{figure*}

Having fixed the parameters of our model, we can extrapolate the amplitudes of elastic $K^-p$
scattering to the subthreshold region, i.e. center-of-mass energies
$1330\,{\rm MeV} \leq W_{\rm cms} \leq 1450\,$MeV. The result is presented in Fig.~\ref{pic:PWA}.
For both real and imaginary parts of the amplitude the maximum lies close
to the $KN$ threshold and is quite narrow, which indicates the presence of a close-by pole. 
It is also worth mentioning that the error band gets smaller to low
energies, different to the recent analysis by Ikeda et al.~\cite{Ikeda:2011pi,Ikeda:2012au}.

To obtain a more complete picture about the structure of $\Lambda(1405)$, 
the amplitudes are analytically continued to the complex $W_{\rm cms}$ plane. 
Microcausality forbids poles on the first Riemann sheet, that is for $\Im(W_{\rm cms})>0$. 
This is fulfilled in our model automatically due to the restoration of analyticity as 
described above. On the other hand some pole structure has to be responsible for 
the functional form of the scattering amplitudes, see  Fig.~\ref{pic:PWA}. 
Two poles are found on the second Riemann sheet for isospin $I=0$, which is 
achieved via analytic continuation to $\Im(W_{\rm cms})<0$. We denote the 
second Riemann sheet connected to the physical axis in the 
region between the $\Sigma\pi$ and $\bar KN$ threshold as
$\mathcal{R}_{\Sigma\pi}$ and the one connected to the physical axis for 
$W_{\rm cms}>(M_K+m_N)$ as $\mathcal{R}_{KN}$. We find that two poles 
lie on different Riemann sheets, the pole position reads
\begin{align*}
\mathcal{R}_{\Sigma\pi}:&~~	W_1=1428_{-1}^{+2}-i~8^{+2}_{-2}~\text{MeV}~.\\
\mathcal{R}_{KN}:&~~		W_2=1467_{-7}^{+11}-i~75^{+9}_{-9}~\text{MeV}~.
\end{align*}
The real part of the position of the first pole agrees quite well with determination from
Refs.~\cite{Borasoy:2006sr,Oller:2000fj,Ikeda:2011pi,Ikeda:2012au}. Its
imaginary part agrees roughly with the determination of 
Refs.~\cite{Borasoy:2006sr,Oller:2000fj} and is significantly smaller than 
extracted by Ikeda et al. ~\cite{Ikeda:2011pi,Ikeda:2012au}. For the second
pole, the situation is different, its imaginary part is in agreement with 
Refs.~\cite{Borasoy:2006sr,Ikeda:2011pi,Ikeda:2012au}, but the real part 
is much larger.

We have investigated the origin of these observations qualitatively. 
First, from the analysis of  $\pi N$ scattering in the 
same framework, see Ref.~\cite{Bruns:2010sv}, it is known that 
off-shell effects  can account for large modifications of  the pole
positions. Setting the tadpole integrals to zero, we obtain immediately the 
solution of the BSE in the on-shell factorization. Note that this solution is still 
different to the one by Ikeda et al.~\cite{Ikeda:2011pi,Ikeda:2012au} since 
no s-wave projection is performed. We found that in the present case the off-shell effects 
do not alter the pole position drastically. More precisely, the imaginary part 
of the first pole decreases and the one of the second increases by about $10$~MeV. The real 
parts of both poles do not change significantly. Secondly, we noticed much 
smaller values of the NLO LECs found by Ikeda et al. additionally to the 
fact that the LECs $b_i$ $(i=5,..,11)$ were neglected there due to the s-wave projection. 
To keep track of this we scale down our LECs continuously from the values 
found above to zero. Such a solution of the BSE is of course by no means 
physical since no further fitting to experimental data is done here. Qualitatively, 
however, we observe that both poles move (the second one by about $100$~MeV) 
to lower values of $\Re(W_{\rm cms})$. The conclusion to be drawn is that 
difference  in pole positions extracted in our approach and the one by 
Ikeda et al. is due to the differences in the fit strategies.

\section*{Acknowledgments}

We are grateful to Peter Bruns for his stimulating remarks and cooperation. We 
thank Michael D\"oring and Wolfram Weise for comments. This work is supported in part by
the DFG (SFB/TR 16 ``Subnuclear Structure of Matter'') and by the EU HadronPhysics3 project
``Study of Strongly Interacting Matter''.

\appendix
\section{Couplings}
\label{app:coupling}
For the channel indices $\{b,j;i,a\}$ corresponding to the process $\phi_iB_a\rightarrow\phi_jB_b$ the relevant coupling matrices read
\begin{align*}
A_{WT}^{b,j;i,a}&=-\frac{1}{4F_j F_i}\langle\lambda^{b\dagger}[[\lambda^{j\dagger},\lambda^{i}],\lambda^{a}]\rangle,\\
A_{14}^{b,j;i,a}&=-\frac{2}{F_j F_i}\Big(
~  2b_4 \langle\lambda^{b\dagger}\lambda^{a}\rangle \langle\lambda^{j\dagger}\lambda^{i}\rangle\\
& + b_1\Big(\langle\lambda^{b\dagger}[\lambda^{j\dagger},[\lambda^{i},\lambda^{a}]]\rangle +\langle\lambda^{b\dagger}[\lambda^{i},[\lambda^{j\dagger},\lambda^{a}]]\rangle\Big)\\
& + b_2\Big(\langle\lambda^{b\dagger}\{\lambda^{j\dagger},[\lambda^{i},\lambda^{a}]\}\rangle +\langle\lambda^{b\dagger}\{\lambda^{i},[\lambda^{j\dagger},\lambda^{a}]\}\rangle\Big)\\
& + b_3\Big(\langle\lambda^{b\dagger}\{\lambda^{j\dagger},\{\lambda^{i},\lambda^{a}\}\}\rangle +\langle\lambda^{b\dagger}\{\lambda^{i},\{\lambda^{j\dagger},\lambda^{a}\}\}\rangle\Big)
\Big),\\
A_{57}^{b,j;i,a}&=-\frac{2}{F_j F_i}\Big(~
b_5\langle\lambda^{b\dagger}[[\lambda^{j\dagger},\lambda^{i}],\lambda^{a}]\rangle+
b_6\langle\lambda^{b\dagger}\{[\lambda^{j\dagger},\lambda^{i}],\lambda^{a}\}\rangle\\
&+
b_7\Big(\langle\lambda^{b\dagger}\lambda^{j\dagger}\rangle \langle\lambda^{i}\lambda^{a}\rangle+
\langle\lambda^{b\dagger}\lambda^{i}\rangle \langle\lambda^{a}\lambda^{j\dagger}\rangle\Big)
\Big),\\
A_{811}^{b,j;i,a}&=-\frac{1}{F_j F_i}\Big(~
2b_{11}\langle\lambda^{b\dagger}\lambda^{a}\rangle \langle\lambda^{j\dagger}\lambda^{i}\rangle\\
&+ b_8~\Big(
\langle\lambda^{b\dagger}[\lambda^{j\dagger},[\lambda^{i},\lambda^{a}]]\rangle 
+\langle\lambda^{b\dagger}[\lambda^{i},[\lambda^{j\dagger},\lambda^{a}]]\rangle
\Big)  \\ 
&+b_9~\Big(
\langle\lambda^{b\dagger}[\lambda^{j\dagger},\{\lambda^{i},\lambda^{a}\}]\rangle 
+\langle\lambda^{b\dagger}[\lambda^{i},\{\lambda^{j\dagger},\lambda^{a}\}]\rangle\Big)\\
& +b_{10}\Big(\langle\lambda^{b\dagger}\{\lambda^{j\dagger},\{\lambda^{i},\lambda^{a}\}\}\rangle 
+\langle\lambda^{b\dagger}\{\lambda^{i},\{\lambda^{j\dagger},\lambda^{a}\}\}\rangle\Big)
\Big),\\
A_{M}^{b,j;i,a}&=-\frac{1}{2 F_j F_i}\Big(~
2b_0\big(
\langle\lambda^{b\dagger}\lambda^{a}\rangle \langle[\lambda^{j\dagger}\lambda^{i}]\bar{\mathcal{M}}\rangle
\big)\\
&+b_D\big(
\langle\lambda^{b\dagger}\{\{\lambda^{j\dagger},\{\bar{\mathcal{M}},\lambda^{i}\}\},\lambda^{a}\}\rangle +
\langle\lambda^{b\dagger}\{\{\lambda^{i},\{\bar{\mathcal{M}},\lambda^{j\dagger}\}\},\lambda^{a}\}\rangle
\big)\\
& +b_F\big(
\langle\lambda^{b\dagger}[\{\lambda^{j\dagger},\{\bar{\mathcal{M}},\lambda^{i}\}\},\lambda^{a}]\rangle+
\langle\lambda^{b\dagger}[\{\lambda^{i},\{\bar{\mathcal{M}},\lambda^{j\dagger}\}\},\lambda^{a}]\rangle
\big) 
\Big),
\end{align*}
where $\lambda$ denote the $3\times 3$ channel matrices (e.g. $\phi =
\phi^{i}\lambda^{i}$ for the physical meson fields) and the $F_i$ are the 
meson decay constants  in the respective channel.
Moreover, $\bar{\mathcal{M}}$ is obtained from the quark mass matrix $\mathcal{M}$
via the Gell--Mann Oakes Renner 
relations, and given in terms of the meson masses as follows,
$\bar{\mathcal{M}}=\frac{1}{2}{\rm diag}(M_{K^+}^2 - M_{K^0}^2 + M_{\pi^0}^2,
M_{K^0}^2 - M_{K^+}^2 + M_{\pi^0}^2, M_{K^+}^2 + M_{K^0}^2 - M_{\pi^0}^2)\,$.


\end{document}